\documentclass[twocolumn,aps,prl,reprint,superscriptaddress,amsmath,amssymb,bbm,floatfix]{revtex4-2}
\usepackage[latin9]{inputenc}
\usepackage{color}
\usepackage{amstext}
\usepackage{graphicx}
\usepackage{esint}
\usepackage[unicode=true,
 bookmarks=false,
 breaklinks=false,pdfborder={0 0 1},colorlinks=true]
 {hyperref}

\usepackage{xcolor,graphicx}
\usepackage{newlfont}
\usepackage{amssymb,amsmath,mathrsfs}
\usepackage{verbatim}
\usepackage{bbm}
\usepackage{multirow}
\usepackage[varg]{txfonts}
 
\newcommand{\be}{\begin{equation}}
\newcommand{\ee}{\end{equation}}
\newcommand{\beq}{\begin{eqnarray}}
\newcommand{\eeq}{\end{eqnarray}}

\newcommand{\tx}{\text}
\newcommand{\mc}{\mathcal}
\newcommand{\ms}{\mathscr}
\newcommand{\mbb}{\mathbbm}

\newcommand{\mf}{\mathfrak}

\newcommand{\ket}[1]{\mbox{$ | #1 \rangle $}}
\newcommand{\bra}[1]{\mbox{$ \langle #1 | $}}

\newcommand{\orcid}[1]{\href{https://orcid.org/#1}{\includegraphics[width=7pt]{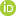}}}

\begin{document}

\title{Experimental assessment of physical realism in a quantum-controlled device}

\author{Pedro R. Dieguez\orcid{0000-0002-8286-2645}}\email{dieguez.pr@gmail.com}
\affiliation{Center for Natural and Human Sciences, Federal University of ABC, Avenida dos Estados 5001, 09210-580, Santo Andr\'{e}, S\~{a}o Paulo, Brazil}
\affiliation{International Centre for Theory of Quantum Technologies (ICTQT), University of Gda\'nsk, Jana Bazynskiego 8, 80-309 Gda\'nsk, Poland}

\author{J\'{e}ferson R. Guimar\~{a}es\orcid{0000-0001-5649-5557}}
\affiliation{Center for Natural and Human Sciences, Federal University of ABC, Avenida dos Estados 5001, 09210-580, Santo Andr\'{e}, S\~{a}o Paulo, Brazil}
\affiliation{Peter Gr\"unberg Institute, Institute for Functional Quantum Systems (PGI-13), Forschungszentrum J\"ulich, North Rhine-Westphalia, 52425 J\"ulich, Germany.}

\author{John P. S. Peterson\orcid{0000-0001-6283-8157}}
\affiliation{Institute for Quantum Computing and Department of Physics and Astronomy,
University of Waterloo, 200 University Avenue West, N2L 3G1, Waterloo, Ontario, Canada}

\author{Renato M. Angelo\orcid{0000-0002-7832-9821}}
\affiliation{Department of Physics, Federal University of Paran\'a, PO BOX 19044, 81531-980, Curitiba, Paran\'a, Brazil} 

\author{Roberto M. Serra\orcid{0000-0001-9490-3697}}\email{serra@ufabc.edu.br}
\affiliation{Center for Natural and Human Sciences, Federal University of ABC, Avenida dos Estados 5001, 09210-580, Santo Andr\'{e}, S\~{a}o Paulo, Brazil}

\begin{abstract} 

Bohr's complementarity principle has been challenged by quantum delayed-choice experiments wherein quantum systems are claimed to behave neither as wave nor as a particle, but in an intermediary way. However, this conclusion has been supported by retro-inference and with no direct link with the system quantum state. Here, we consider a framework that employs an operational criterion of physical reality to diagnosis the system ontology directly from the quantum state at each instant of time. We show that, in disparity with previous proposals, our setup ensures a formal link between the output visibility and elements of reality within the interferometer. An experimental proof-of-principle is provided for a two-spin-1/2 system in an interferometric setup implemented in a nuclear magnetic resonance platform. We discuss how our results validate, to a great extent, Bohr's original formulation of the complementarity principle and unveil morphing reality states.
\end{abstract}

\maketitle


Although lacking an indisputable formulation in terms of the mathematical structure of quantum mechanics, Bohr's complementarity principle~\cite{Bohr35} manifested its pivotal role in modern physics by submitting matter and radiation to an unifying framework: any one of these elements is expected to exclusively behave either in a  wave or in a  particle-like manner, depending on the peculiarities of the experimental setup. Bohr's natural philosophy~\cite{Saunders2005} also advocates the impossibility of ascribing individuality to quantum systems, meaning that the physical reality cannot emerge until the whole experiment, including the system and the classical measuring apparatus, is definitely arranged. These ideas prompted Wheeler to devise the concept of a delayed-choice experiment~\cite{Wheeler}, a scenario wherein the classical apparatus, typically an interferometer, is settled only when the quantum system has entered it. When this experiment came into actual existence~\cite{Jackes07,Manning15,Vedovato20}, the observed visibility at the output of the interferometer always revealed the fingerprints of wave or particle in full agreement with the results expected for the (posteriorly defined) corresponding setting. The complementarity principle was not cheated by the delayed arrangement of ``the whole''. Moreover, generalizations of Wheeler's idea in terms of entanglement-separability duality of bipartite systems were also proposed~\cite{Peres00,Jennewein05,Bruckner05} and experimentally confirmed in the context of delayed-choice entanglement swapping experiments~\cite{Sciarrino02,Ma12}.

Nearly one decade ago, a quantum delayed-choice experiment (QDCE) was conceived~\cite{Ionicioiu11} in which one beam-splitter is prepared in a spatial quantum superposition, thus rendering the interferometer to have a ``closed $+$ open'' configuration and the system to be in a hybrid ``wave $+$ particle'' state. To test these ideas, researchers coupled a target system to a  quantum controller---an ancilla in superposition that effectively implements a suspended configuration for the interferometer~\cite{Auccaise12,Roy2012,Peruzzo12,Kaiser12}. Besides allowing for the control of the superposition degree of the whole system, such strategy enables one to decide (measure) the configuration after the target has traversed the interferometer. The capability of such scheme to smoothly interpolate the observed statistics between a wave- and a particle-like pattern suggested the manifestation of ``morphing behaviours'' in the same setup, thus claiming for a radical revision of Bohr's original statement of the complementarity principle~\cite{Adesso12}. More recently, variants of this experiment have reported on the observation of an entangled duality in a two-photon system~\cite{Rab17}, a which-path detector that can simultaneously record and neglect the system's path information~\cite{Liu17}, and the implementation of a nonlocal setting control~\cite{Wang2019}. Entropic frameworks have offered alternative interpretations for the QDCE~\cite{Coles14,Angelo15} and the equivalence of such setting  with a prepare-and-measure scenario in the perspective of device independent causal models has been demonstrated~\cite{Chaves18}. 

A common feature of all these experiments is the use of retro-inference  about the system behaviour inside the interferometer with basis on the visibility observed at the output of the interferometer. To date, a detailed analysis is lacking which would allow one to track the behavior of the system at every stage of the experiment. Moreover, it is still not clear how elements of reality emerge from ``the whole'' and whether quantum correlations play some fundamental role. In this work, these questions are thoroughly addressed. First, we adopt an operational quantifier of realism that explicitly depends on the quantum state and allows for meaningful which-path statements. This enables us to discuss realism for ``the whole'' by looking at the global state of the system at each instant of time. Second, we show that the visibility at the output has no connection whatsoever with wave and particle elements of reality, as defined in accordance with the adopted criterion of realism. This raises important objections to the usual interpretations of the QDCE. Third, we propose a setup that removes these objections and establishes a monotonic link between the visibility and wave elements of reality inside the interferometer. We then demonstrate the relevance of quantum correlations to wave-particle duality. Fourth, by use of a nuclear magnetic resonance (NMR) platform, we submit our model to experimental scrutiny. Finally, we argue how our results retrieve Bohr's original view of the complementarity principle.
%

%

{\bf \large{Results}}

\textbf{Contextual realism in the QDCE.}

Recently, a criterion of realism has been put forward~\cite{Bilobran15} with basis on a single premise, namely, that after a projective measurement is performed of a physical quantity $\ms{A}$, represented by a discrete-spectrum observable $A=\sum_aaA_a$, with projectors $A_a=\ket{a}\bra{a}$ acting on $\mc{H_A}$, for a given preparation $\varrho$ on $\mc{H_A\otimes H_B}$, then $\ms{A}$ becomes an element of reality, even if the measurement outcome is not revealed. Accordingly, the post-measurement state $\Phi_A(\varrho)\coloneqq\sum_a (A_a\otimes\mbb{1})\varrho(A_a\otimes \mbb{1})$ is taken as a primitive notion of $A$-reality state. It then follows that
\be
\label{irrealism}
\mf{I}_A(\rho)\coloneqq \min_{\varrho} S\big(\rho||\Phi_A(\varrho)\big)=S\big(\Phi_A(\rho)\big)-S(\rho)
\ee
is a faithful quantifier of $A$-realism violations for a given state $\rho$ (where $S(\rho):=-\text{Tr}\big(\rho\log_2{\rho}\big)$ is the von Neumann entropy). By virtue of the properties of the relative entropy of $\rho$ and $\sigma$,  $S(\rho||\sigma)=\text{Tr}\big[\rho(\log_2{\rho}-\log_2{\sigma})\big]$, the so-called  irrealism of the context $\{A,\rho\}$ is bounded as $0\leq\mf{I}_A(\rho)\leq\log_2{d_\mc{A}}$, with $d_\mc{A}=\dim\mc{H_A}$, vanishing iff $\rho=\Phi_A(\rho)$. This measure has been applied to a number of foundational investigations~\cite{Dieguez18,Rudnicki18,Gomes18,Gomes19,Fucci19,Orthey19,Engelbert20,Lustosa20}, including an experimental test in a photonic platform~\cite{Mancino2018}. Also, irrealism has formally been framed as a quantum resource~\cite{Costa2020}. Two properties of irrealism will be crucial here. First, since $\mf{I}_A(\rho)-\mf{I}_A(\rho_\mc{A})\geq \mc{D_A}(\rho)$~\cite{Bilobran15}, where $\rho_\mc{A}=\text{Tr}_\mc{B}(\rho)$ is the reduced state, $\mc{D_A}(\rho)=min_{A}[I_\mc{A:B}(\rho)-I_\mc{A:B}(\Phi_A(\rho))]$ is the quantum discord~\cite{Ollivier01,Henderson01,Celeri11,Modi12}, and $I_\mc{A:B}(\rho)=S(\rho||\rho_\mc{A}\otimes\rho_\mc{B})$ denotes the mutual information, one has that, whenever quantum discord  is present, the $A$-irrealism induced by the joint state $\rho$ is greater than the one deriving from the part $\mc{A}$ alone. This shows that quantum correlations render irrealism to be a property of ``the whole''. Second, the relation $\mf{I}_A(\rho)+\mf{I}_{A'}(\rho)\geq S\left(\rho||\frac{\mbb{1}}{d_A}\otimes\rho_\mc{B}\right)$~\cite{Freire19} concerning maximally incompatible observables $A$ and $A'$ acting on $\mc{H_A}$ precludes the manifestation of full realism whenever $\rho\neq \frac{\mbb{1}}{d_A}\otimes\rho_\mc{B}$. For convenience, here we follow the ideas discussed in Ref.~\cite{Dieguez18} to work instead with the $A$-realism
\be
\label{realism}
\mf{R}_A(\rho):=\log_2{d_{\cal{A}}}-\mf{I}_{A}(\rho),
\ee
which quantifies how close the scenario is to the $A$-realistic context $\{\Phi_A(\rho),A\}$. It follows from the above relations that $\mf{R}_A(\rho_\mc{A})-\mf{R}_A(\rho)\geq \mc{D_A}$ (non-separability of $A$-realism) and
\be
\label{CP}
\mf{R}_A(\rho)+\mf{R}_{A'}(\rho)\leq\log_2{d_\mc{A}}+S(\rho_\mc{A})-I_\mc{A:B}(\rho).
\ee
Notably, these correlations further restricts quantum systems to reach full realism. For pure states $\rho=\ket{\psi}\bra{\psi}$, the upper bound becomes $\log_2{d_\mc{A}}-E(\psi)$, with $E(\psi)=S(\rho_\mc{A(B)})$ the entanglement entropy of $\ket{\psi}$.

\begin{figure}[t]
\centering
\includegraphics[width=8.25cm]{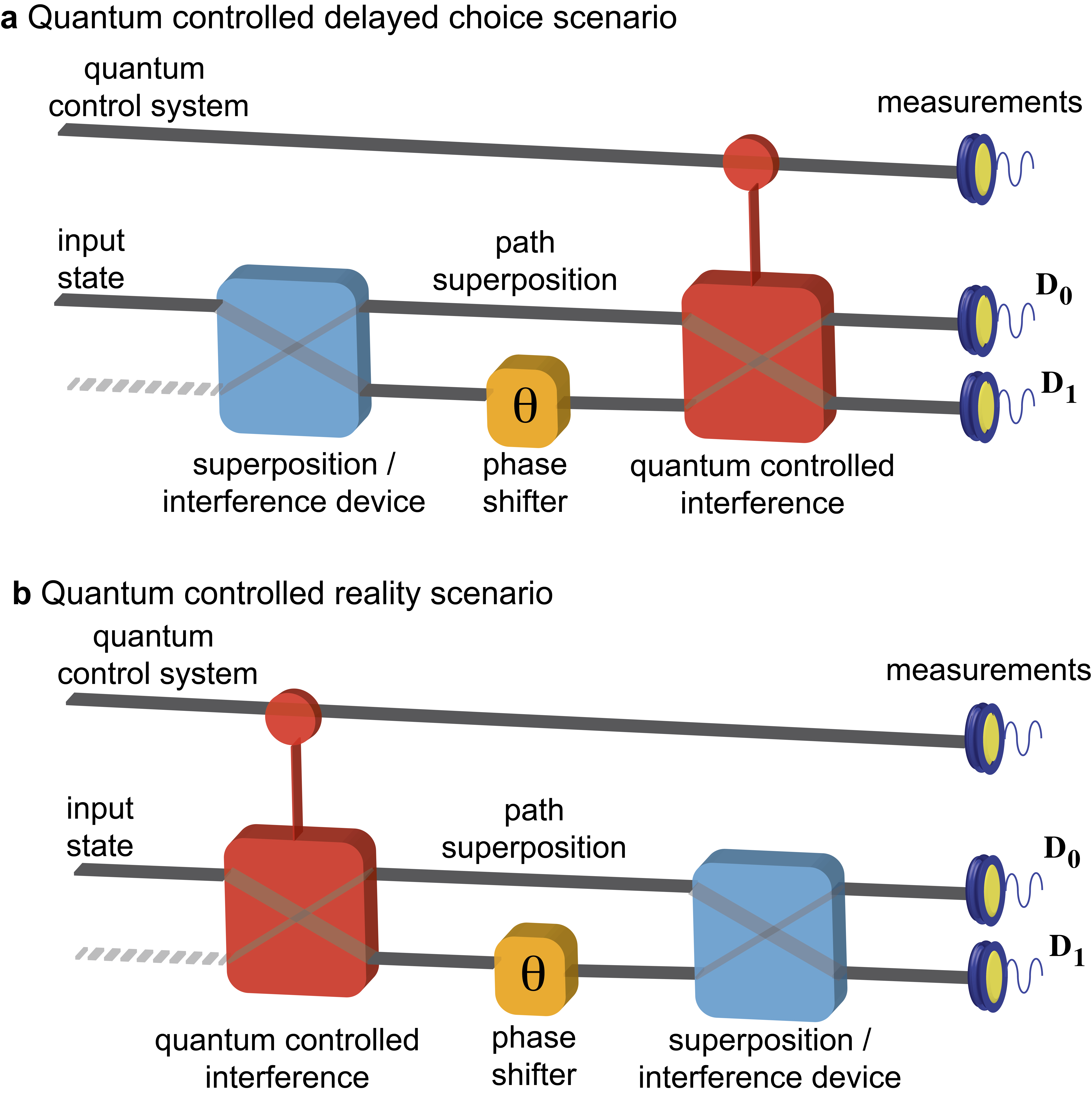}
\caption{\textbf{Schematic circuits of quantum controlled interferometers}. The blue boxes represent unitary operations which here play the role of superposition devices---the quantum network equivalent of a beam-splitter. Using an ancillary qubit in superposition (quantum control system), we implement the quantumly controlled unitary superposition device (represented by the red boxes). {\bf a} Original version of the quantum delayed-choice experiment, where the second beam-splitter is prepared in a coherent superposition of being {\it in} and {\it out} of the interferometer (configurations {\it closed} and {\it open}, respectively). {\bf b} Our proposal for a quantum controlled reality experiment. Here, the first beam-splitter is submitted to quantum control. Although the measurement outcomes yield the same visibility in both of these experimental arrangements, the realism aspects inside the interferometer are crucially different.} 
\label{Fig1}
\end{figure}

Equipped with the above tools, we now reassess the QDCE (depicted in the first circuit in Fig.~\ref{Fig1}{\bf a}) from the perspective of elements of reality. As discussed in the original formulation~\cite{Ionicioiu11}, a qubit is put in a what was previously pictured as a  particle-like state $\ket{\wp_\theta}=\frac{1}{\sqrt{2}}\left(e^{i\theta}\ket{0}+\ket{1}\right)$ after passing the first superposition device (or beam-spliter) and the phase shifter, which implements a relative phase between the paths $\{0,1\}$ travelled by the qubit.  When the final superposition device is activated in Fig.~\ref{Fig1}\textbf{a} (interferometer closed) in a typical Mach-Zender interferometer, then the state $\ket{\wp_\theta}$ transforms into what is considered as a  wave-like state $\ket{w_\theta}=e^{i\theta/2}\left(\cos{\frac{\theta}{2}} \ket{0}+i\sin{\frac{\theta}{2}}\ket{1}\right)$ (according to the picture introduced in Ref.~\cite{Ionicioiu11}), with amplitudes clearly depending on the phase $\theta$. The QDCE is realized by preparing the quantum controller ($\mc{C}$) in a general coherent superposition of being closed (in) and open (out), {\it i.e.}, $\cos{\frac{\alpha}{2}}\ket{\tx{in}}+\sin{\frac{\alpha}{2}}\ket{\tx{out}}$,  with $0\leq\alpha\leq\pi$ (see Methods for more details on the experimental preparation of the initial state). In other words, $\alpha$ determines the probabilities of the second beam-splitter operation to be found ``in'' the circuit (defining abstractly as the state $\ket{\tx{in}}$) or ``out'' of it ($\ket{\tx{out}}$), as depicted in Fig.~\ref{Fig1}\textbf{a}. According with this original picture, the result of such interaction at the output of the interferometer reads
$\ket{\psi_\tx{\bf a}}=\cos{\frac{\alpha}{2}}\ket{w_\theta}\ket{\tx{in}}+\sin{\frac{\alpha}{2}}\ket{\wp_\theta}\ket{\tx{out}}$. The interference pattern in the detector $D_0$ (of Fig.~\ref{Fig1}\textbf{a}) is then shown to be written as $\mathfrak{p}_0=\text{Tr}(\ket{0}\bra{0}\otimes\mbb{1}_\mc{C} \,\rho_\tx{\bf a})=\frac{1}{2}\left(1+\mc{V}\cos{\theta}\right)$, where $\rho_\tx{\bf a}=\ket{\psi_\tx{\bf a}}\bra{\psi_\tx{\bf a}}$ and $\mc{V}\coloneqq (\mathfrak{p}_0^\tx{max}-\mathfrak{p}_0^\tx{min})/(\mathfrak{p}_0^\tx{max}+\mathfrak{p}_0^\tx{min})=\cos^2\left(\frac{\alpha}{2}\right)$, which stands for the visibility of the interference pattern computed from optimizations running over $\theta$. The association of $\ket{w_\theta}$ and $\ket{\wp_\theta}$ with wave- and particle-like behaviours is justified with basis on the resulting dependence of $\mathfrak{p}_0$ on the phase $\theta$, as can be readily checked for $\alpha=0$ and $\alpha=\pi$, respectively. Moreover, the wave-like amplitude in $\ket{\psi_\tx{\bf a}}$ is clearly related to the visibility $\mc{V}$. The scenario is such that by looking at the statistics at the output of the circuit, one infers the way the qubit travelled the interferometer. However natural this argument may sound, it suffers from an important flaw: regardless of how high $\mc{V}$ may be, thus presuming an accentuated wave-like behaviour, the qubit state inside the interferometer invariably is $\ket{\wp_\theta}$, which has been assumed to be linked with particle-like behaviour. If, on the other hand, we argue that this state cannot be used to account for the qubit route, then we are somehow conceiving that quantum mechanics is not complete. To further stress the issue, we compute realism inside the circuit. 

We propose here a framework to discuss the elements of reality (for the wave-particle behaviour) in a quantum-controlled interference device. Let us now define $P\equiv\sigma_z$ and $W\equiv\sigma_{\perp}$, with respective eigenstates $\{\ket{\mc{P}_+},\ket{\mc{P}_-}\}\equiv\{\ket{0},\ket{1}\}$ and $\ket{\mc{W}_\pm}=\frac{1}{\sqrt{2}}\left(e^{i\theta}\ket{0}\pm\ket{1}\right)$, as the relevant eigenstates for the particle and wave observables, respectively. This choice naturally connects definite paths, $\ket{\mc{P}_\pm}$, with particle-like elements of reality. Formally, one sees that $\Phi_P(\mc{P}_\pm)=\ket{\mc{P}_\pm}\bra{\mc{P}_\pm}$, showing that eventual measurements of $P$ would not change the state of affairs. Hence, $P$ is an already installed element of reality. Accordingly, with Eqs.~\eqref{irrealism} and \eqref{realism} we find $\mf{R}_P(\mc{P}_\pm)=1$ and $\mf{R}_W(\mc{P}_\pm)=0$. On the other hand, since $\Phi_P(\mc{W}_\pm)\neq \ket{\mc{W}_\pm}\bra{\mc{W}_\pm}$, the states $\ket{\mc{W}_\pm}$ cannot be related to a path element of reality. In fact, superposed paths imply wave-like elements of reality (wave reality, for short), since $\Phi_W(\mc{W}_\pm)=\ket{\mc{W}_\pm}\bra{\mc{W}_\pm}$ and then $\mf{R}_W(\mc{W}_\pm)=1$ and $\mf{R}_P(\mc{W}_\pm)=0$. In this perspective, one readily finds $\ket{\wp_\theta}=\ket{\mc{W}_+}$ and
\be
\mf{R}_P\left(\wp_\theta\right)=0\qquad\tx{and}\qquad\mf{R}_W\left(\wp_\theta\right)=1,
\ee
showing that the so-called particle-like state actually corresponds to a wave reality. An important physical argument corroborating this view is that $\ket{\wp_\theta}$ encapsulates a detectable relative phase in the basis $\{\ket{\mc{P}_\pm}\}$ and can be used to create entanglement between far distant particles \cite{Angelo15}. Therefore, in full contrast to current claims, inside the interferometer (in the QDCE described in Fig.~\ref{Fig1}\textbf{a}) the qubit always behaves as a wave and $\mc{V}$ does not faithfully furnish this diagnosis. It follows that the pattern $\mathfrak{p}_0$, for $0<\alpha<\pi$, cannot validate the manifestation of wave and particle morphing behaviour in the setup of Fig. \ref{Fig1}{\bf a}. Finally, it should be noticed that our approach conceives the physical reality as being determined solely by the quantum state at every instant of time. With that we reject  retrocausal models according to which measuring the visibility at the output may determine the state of affairs inside the interferometer (see Methods for further discussions).

\begin{figure}[t]
\centering
\includegraphics[width=8.10cm]{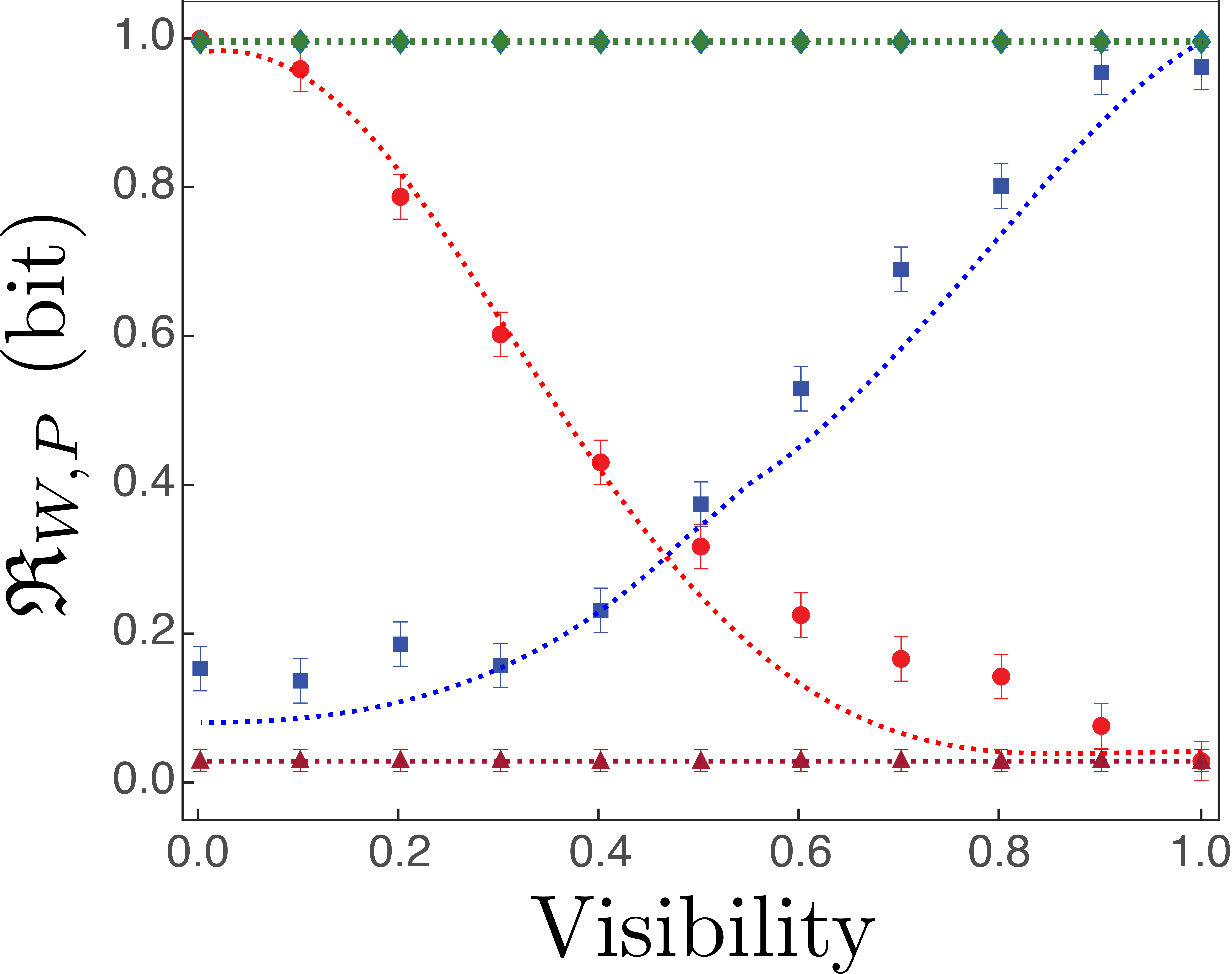}
\caption{\textbf{Wave and Particle Realism as a function of the Visibility}. The green diamonds and dark red triangles are the measured $\mf{R}_W$ (wave realism) and $\mf{R}_P$ (particle realism), respectively, inside of the interferometer with the arrangement in Fig. \ref{Fig1}{\bf a} (quantum delayed-choice experiment). The blue squares and red circles are the measured $\mf{R}_W$ and $\mf{R}_P$, respectively, inside of the interferometer of Fig. \ref{Fig1}{\bf b} (quantum-controlled reality experiment). The symbols represent the experimental results and the dashed lines are numerical calculations which simulate the pulse sequences on the initial experimental state. The data is parametrized by the visibility at the end of the interferometer. The error bars were estimated via Monte Carlo propagation (see Error Analysis subsection in Methods for further details). The error bars for data represented as green diamonds are smaller than the symbols.  
}
\label{Fig2}
\end{figure}

\begin{figure*}[t]
\centering
\includegraphics[width=17cm]{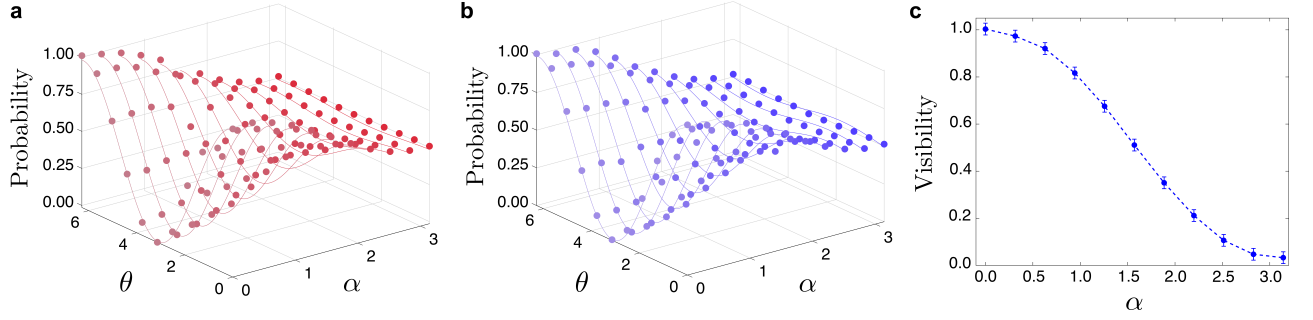}
\caption{\textbf{Probability pattern at the end of the interferometer ($\mathfrak{p}_0$) as function of the interference parameter ($\alpha$) and the phase shifter ($\theta$)}. {\bf a} For quantum controlled delayed choice scenario (setup in Fig. \ref{Fig1}{\bf a}). {\bf b} For quantum controlled realism scenario (setup in Fig. \ref{Fig1}{\bf b}). {\bf c} Visibility ($\mc{V}$) of the interferometer in the quantum controlled realism scenario. The symbols represent the experimental results and the (solid and dashed) lines numerical simulations. The  error bars were estimated via Monte Carlo propagation (see Error Analysis subsection in Methods for further details). In panels \textbf{a} and \textbf{b}, the error bar is smaller than the symbols.}
\label{Fig3}
\end{figure*}

\textbf{ Quantum-controlled reality experiment (QCRE)}

We now propose an experiment that aims at solving the aforementioned issues and effectively superposing wave and particle elements of reality. Consider the setup depicted in Fig. \ref{Fig1}{\bf b}, which implements a simple exchange in the devices order. The input state $\ket{\psi_{in}}=\ket{0}\left(\cos{\frac{\alpha}{2}}\ket{\tx{in}}+\sin{\frac{\alpha}{2}}\ket{\tx{out}}\right)$ now results in  
$\ket{\psi_\tx{\bf b}}=\cos{\frac{\alpha}{2}}\ket{w_\theta}\ket{\tx{in}}+e^{i\theta}\sin{\frac{\alpha}{2}}\ket{\wp_0}\ket{\tx{out}}$, thus yielding precisely the same interference pattern
$\mathfrak{p}_0=\frac{1}{2}\left(1+\mc{V}\cos{\theta}\right)$ and visibility $\mc{V} = \cos^2\left(\frac{\alpha}{2}\right)$. Clearly, with respect to the statistics observed at the output, nothing changes. However, the states of the whole system when the qubit is traveling inside the interferometer right after the phase shifter are, for the two scenarios of Fig.~\ref{Fig1}, given by
\begin{subequations}
\beq 
&&\ket{\phi_\tx{\tiny QDCE}}=\ket{\mc{W}_+}\left(\cos{\tfrac{\alpha}{2}}\ket{\tx{in}}+\sin{\tfrac{\alpha}{2}}\ket{\tx{out}}\right), \label{psi-QDCE}\\ 
&&\ket{\phi_\tx{\tiny QCRE}}
=\cos{\tfrac{\alpha}{2}}\ket{\mc{W}_+}\ket{\tx{in}}+e^{i\theta}\sin{\tfrac{\alpha}{2}}\ket{\mc{P}_+}\ket{\tx{out}}. \label{phi-QCRE}
\eeq 
\end{subequations}
The differences are remarkable. Correlations are seen to play no role in the original setup (QDCE described in Fig.~\ref{Fig1}\textbf{a}), so that there is not an effective ``whole'' defining the behavior of the qubit at this stage. The interference device puts the qubit in a superposition of paths, which implies a wave reality: 
\be \label{R-QDCE}
\mf{R}_W(\phi_\tx{\tiny QDCE})=1,\quad\quad \mf{R}_P(\phi_\tx{\tiny QDCE})=0. 
\ee
In the new setup (QCRE described in Fig.~\ref{Fig1}\textbf{b}), when the controlled interference device is deactivated, the qubit just keeps travelling its original path, as a particle, this being a key difference to the original QDCE. The state \eqref{phi-QCRE} predicts a clear superposition of two orthogonal reality scenarios: one wave-like $(\ket{\mc{W}_+}\ket{\tx{in}})$ and another particle-like $(\ket{\mc{P}_+}\ket{\tx{out}})$. In addition, via direct calculations one finds
\be \label{R-QCRE}
\mf{R}_W(\phi_\tx{\tiny QCRE})=1-h\left(\tfrac{1-\mc{V}}{2}\right),\quad\quad\mf{R}_P(\phi_\tx{\tiny QCRE})=1-h\left(\tfrac{\mc{V}}{2}\right),
\ee
where $h(u)=-u\log_2{u}-(1-u)\log_2{(1-u)}$ is the binary entropy. These relations show that $\mf{R}_W$ ($\mf{R}_P$) is a monotonically increasing (decreasing) function of the visibility $\mc{V}$. Hence, in contrast to the QDCE, here we have a strict equivalence between the output statistics and the wave-like behavior inside the interferometer. Note that within the domain $\alpha\in[0,\pi]$, the visibility $\mc{V}$ at the output and the wave and particle realisms $\mf{R}_{W,P}$ inside the interferometer are monotonic functions of $\alpha$, meaning that these quantities can be controlled by the preparation of the quantum controller $\mc{C}$. Also, relation \eqref{CP} here reduces to
\be \label{CP-QCRE}
\mf{R}_W(\phi_\tx{\tiny QCRE})+\mf{R}_P(\phi_\tx{\tiny QCRE})\leq 1-h\left(\frac{1+\lambda_\mc{V}}{2}\right),
\ee 
where $\lambda_\mc{V}\equiv\tx{\small $\sqrt{2\mc{V}^2-2\mc{V}+1}$}$. This result as a function of $\mc{V}$ demonstrates how quantum correlations between the qubit and the quantum controller are sufficient to deny classical realism inside the interferometer to both complementary observables at the same time, thus corroborating Bohr's original formulation of the complementarity principle.

We implemented the above ideas in a prof-of-principle experiment using a liquid-state NMR setup with two spin-1/2 qubits encoded in a sample of $^{13}$C-labelled CHCl$_3$ (Chloroform) diluted in Acetone-d6. The experiments were carried out in a Varian 500 MHz spectrometer. The $^{13}$C nuclear spin was used as the ancillary control of the interferometric device to investigate the realism, wave and particle features of the $^1$H nuclear spin (which encompass the interferometric paths). The nuclear spins where initially prepared in a state equivalent to $\rho = \ket{\psi_{in}}\bra{\psi_{in}}$, using spatial averages techniques~\cite{oliveira2007} (see  Methods).     

All the spin-1/2 quantum controlled interferometric protocols in Fig. \ref{Fig1} were performed using combinations of transverse radio-frequency pulses on resonance with each nuclei and sequences of free evolution under the spin scalar coupling, $H_J=J\frac{h}{4}\sigma_Z^{\text{H}}\otimes \sigma_Z^{\text{C}}$, with $\sigma_Z^{\text{H(C)}}$ being the Pauli operator for the $^1$H ($^{13}$C) nuclear spin and $J \approx 215.1$~Hz the coupling constant. The realism (displayed in Fig. \ref{Fig2}) is quantified performing full quantum state tomography~\cite{oliveira2007} along the interferometric protocol with different values of the interference control parameter $\alpha$ and the phase sifter $\theta$ for each setup in Fig. \ref{Fig1}. The interferometric pattern ($\mathfrak{p}_0$) is observed directly from the $^1$H nuclear spin magnetization in the $z$-direction at the end of the protocols.

{\bf \large{Discussion}}

Wave and particle are classical physics terms employed for one to discuss the behaviour of a quantum system that traverses a double-path setup and produces some signals and statistics in the output measurement device. When the signal pattern depends on the difference of phases between the paths, the behavior is claimed to be wavelike. However, in the QDCE the output visibility does not tell an unambiguous story about the qubit behaviour inside the circuit. The state inside the interferometer---the so-called particle state, $\ket{\wp_\theta}$---has spatial coherences, a resource that can be used to create entanglement between distinct quantum systems~\cite{Angelo15}. It is hard to imagine that a presumably well-localized system (a particle), developing a (hidden) realistic trajectory in space-time, be able to touch space-like separated systems. The strategy of making inferences about past behaviour (inside the circuit) with basis on present observations (at the output) diminishes the significance of the quantum state and favours the view according to which quantum mechanics is not a complete theory. In addition, the statistics-based criterion does not shed light on another challenging aspect of typical double-path experiments: even when the observed pattern is wavelike, in each run of the experiment only one of the detectors is activated (this point is discussed in detail in the Methods.)

The realism-based framework we developed here makes a rather different narrative for the QDCE and introduces a QCRE, an arrangement that has the same output visibility of the former, as demonstrated in Fig. \ref{Fig3}, but whereby Bohr's original formulation of the complementarity principle can be afforded, as we show next, a deeper and broader significance. As theoretically predicted [Eq.~\eqref{R-QDCE}] and experimentally corroborated (Fig. \ref{Fig2} and Fig. \ref{Fig3}), wave and particle elements of reality inside the circuit are fully disconnected from the output visibility, so that no wave-particle superposition or morphing behavior can actually be claimed in the QDCE. The scenario is quite the opposite in the QCRE, where wave and particle elements of reality are regulated by the initial coherence of the quantum controller and turn out to be monotonically linked with he visibility, as demonstrated by Eq.~\eqref{R-QCRE} and Fig. \ref{Fig2}. Equation \eqref{phi-QCRE} and our experimental demonstration arguably show, for the first time (to the best of our knowledge), the possibility of genuinely superposing wave and particle elements of reality to an arbitrary degree. By employing the figures of merit $\mf{R}_{W,P}(\rho)$, which lies solely on the time-local context defined by the composite state $\rho$ and observables $\{W,P\}$, thus respecting premises of standard quantum mechanics, our model avoids retro-causal inferences and suitably describe ``the whole''.

Most interestingly, the QCRE allows for the manifestation of ``morphing realities'' (for $0<\mc{V}<1$), but the inequality \eqref{CP-QCRE} prevents $W$ and $P$ to be simultaneous elements of reality. In fact, referring back to the general scenario underlying inequality \eqref{CP}, one can check that the upper bound can be written as $2\log_2{d_\mc{A}}-S\left(\rho||\frac{\mbb{1}}{d_\mc{A}}\otimes\rho_\mc{B}\right)$. This demonstrates that $\rho=\frac{\mbb{1}}{d_\mc{A}}\otimes\rho_\mc{B}$ is the only state admitting the saturation $\mf{R}_A=\mf{R}_{A'}=\log_2{d_\mc{A}}$, for all $A$ and $A'$, in which case full realism emerges in the part $\mc{A}$ (classical regime). For any other (non-classical) state, the mutual exclusiveness of realism is implied. To further appreciate this point, it is instructive to use the conditional entropy $S_\mc{A|B}(\rho)=S(\rho)-S(\rho_B)$ and the conditional information $I_\mc{A|B}(\rho)=\log_2{d_\mc{A}}-S_\mc{A|B}(\rho)$ to rewrite inequality \eqref{CP} in the form
\be \label{BohrCP}
\mf{R}_A(\rho)+\mf{R}_{A'}(\rho)\leq 2\log_2{d_{A}}-I_\mc{A|B}(\rho).
\ee 
Also, it is noteworthy that $I_\mc{A|B}(\rho)=I(\rho_\mc{A})+I_\mc{A:B}(\rho)$, with $I(\rho_\mc{A})=\log_2{d_\mc{A}}-S(\rho_\mc{A})$ being a purity measure for the reduced state. These relations help us to make a fundamental point: the purity of a system and the correlations shared with ``the whole'' prevent it to have simultaneous elements of reality. In other words, provided we consider a quantum state for a composite system (``the whole'') in a given setting, even in a quantumly controlled one, wave and particle realities can never manifest themselves simultaneously. Such statements stand out in full opposition to the Einstein, Podolsky, and Rosen (EPR) arguments about physical reality~\cite{EPR35} and validate Bohr's philosophy. Finally, we highlight that our work sheds light on the role of the complementarity principle in the context of morphing reality states submitted to a quantum controlled operation, which may lead to new insights regarding the nature of quantum causality, quantum reference frames, and, specially, a renewal of the discussion on the realistic aspects of wave and particle properties linked to quantum systems.

{\bf \large{Methods}}
\label{sup}

{\bf Experimental Setup}

In the reported experiments we used a liquid sample of $^{13}$C-labelled CHCl$_3$ (Chloroform) diluted in Acetone-d6, the Chloroform molecules in the sample were composed of four nuclei (isotopes): $^{1}$H, $^{13}$C, $^{35}$Cl, and $^{37}$Cl. We only controlled the $^{1}$H and $^{13}$C nuclei. The sample is very diluted ($\approx 1\%$) in Acetone-d6, such that the Chloroform inter-molecular interactions can be neglected and the sample can be considered as a set of identically prepared pairs of spin-1/2 systems (multiple copies of a two-qubit system). The Chlorine only provides mild environmental effects in the experiments. 

The experiments were performed using a Varian 500 MHz spectrometer equipped with a superconducting magnet, double-resonance probe head with magnetic field-gradient coil. The superconducting magnet produces a static field with B$_{0}$ $\approx$ 11.75~T of intensity, along the positive z-axes (parallel to the central axis of the magnet). Under the action of this static field, the resonance frequencies of $^{1}$H and $^{13}$C nuclei are approximately 500~MHz and 125~MHz, respectively. The state of the nuclear spins are controlled by time-modulated radio frequencies pulses (rf-pulses) in the transverse direction  (x and y), longitudinal field gradients, as well as by sequences of free evolution of the system under the action of scalar coupling interactions. The latter is described by the Hamiltonian $H_J=J\frac{h}{4}\sigma_Z^{\text{H}}\otimes \sigma_Z^{\text{C}}$, where $J \approx 215.1$~Hz is the coupling constant between the $^{1}$H and $^{13}$C nuclear spins.

Measured by inversion recovery, the spin-lattice relaxation times were found to be $T_1 ^H=8.882$~s and $T_1 ^C=18.370$~s. In the case of transverse relaxations, the characteristics times were obtained by the Carr-Purcell-Meiboom-Gill (CPMG) pulse sequence, resulting in $T_2 ^H=2.185$~s and $T_2 ^C=0.310$~s. The interferometric protocols in the experiments were performed in a time scale of $\approx 14$ ms, which is sufficiently smaller than the aforementioned decoherence times for neglecting its effects. 

\begin{figure}[h!]
\centering
\includegraphics[width=8.10cm]{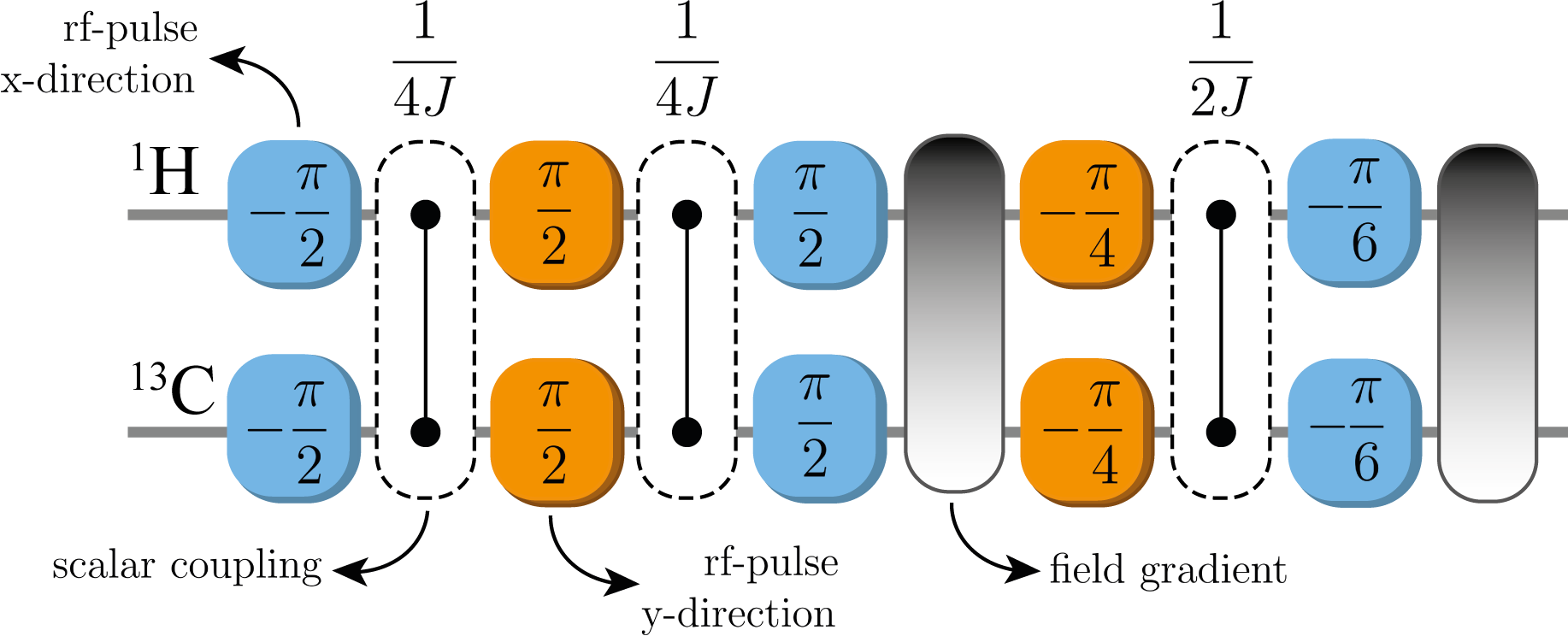}
\caption{\textbf{Pulse sequence for the initial state preparation}. The blue (orange) boxes represent x (y) local rotations by the angles indicated inside. These rotations are produced by a transverse rf-field resonant with either $^{1}$H or $^{13}$C nuclei, with phase, amplitude, and time duration properly adjusted. The black dashed boxes with connections represent free time evolution under the scalar coupling of both spins. The boxes with a gray gradient represent magnetic field gradients, with longitudinal orientations aligned with the spectrometer cylindrical symmetry axis. All the control parameters are optimized to build an initial pseudo-pure state equivalent to $\rho = \ket{00}\bra{00}$ with high fidelity ($\gtrapprox 0.99$).} 
\label{Fig4}
\end{figure}

The effective initial state of the nuclear spins were prepared by spatial average techniques~\cite{oliveira2007,batalhao14,batalhao15}, being the $^{1}$H and $^{13}$C nuclei used as the system of interest and ancillary control, respectively. For both protocols shown in Fig. \ref{Fig1}, the initial pseudo-pure state~\cite{oliveira2007,knill1998effective} equivalent to $\rho = \ket{\psi_{in}}\bra{\psi_{in}}$ was prepared by the pulse sequence depicted in Fig. \ref{Fig4}.

\begin{figure}[h!]
\centering
\includegraphics[width=8.10cm]{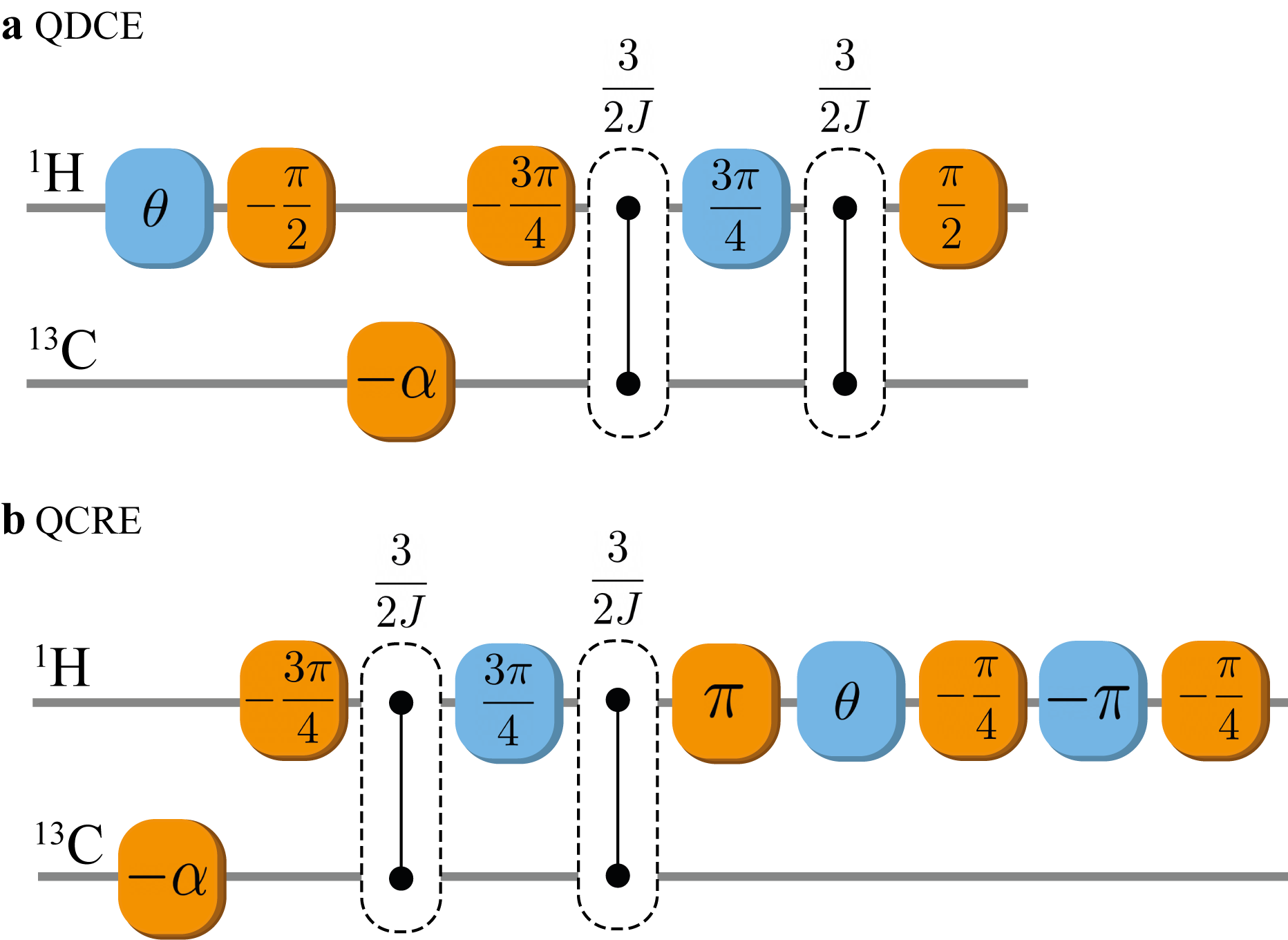}
\caption{\textbf{Pulse sequences for the two interferometric scenarios}. {\bf a} Sequence for the original version of quantum delayed-choice experiment (QDCE) described in Fig.~\ref{Fig1}\textbf{a}. For the sake of optimization, the first superposition operation and the phase shifter were implemented by two rotations (rotations $\theta$ and $-\frac{\pi}{2}$). The quantum-controlled interference was performed using local operations on the system ($^1$H) and on the controller ($^{13}$C), as well as two free evolution under the scalar coupling. {\bf b} Pulse sequence for the quantum-controlled reality experiment (QCRE) described in Fig.~\ref{Fig1}\textbf{b}, where the quantum-controlled interference appears as the first operation followed by the phase shifter and the interference operation. The most relevant contributions to the total time duration of each experiment is the free evolution, so both pulse sequences last approximately the same time ($\approx 14$ ms).} 
\label{Fig5}
\end{figure}

To implement the two quantum-controlled interferometers described in Fig.~\ref{Fig1} in NMR, we designed optimized pulse sequences in order to minimize experimental errors. The correspondent pulse sequence to each interferometric scenario (QDCE and QCRE) are presented in Fig. \ref{Fig5}. In the present experiment, rotations on each nuclear spin are implemented through hard (square) rf-pulses in the transverse direction resonant with the respective nuclei. The pulse generated by a coil in the probe has it's phase adjusted to produce a rotation in the $x$- or $y$-directions. The rotation angle can be set by the time duration of the pulse or the transverse rf-field intensity \cite{oliveira2007,Levitt2008}. In the rotations involving the phase shifter $\theta$ and the controlled-interference parameter $\alpha$, we linearly varied the rf-field intensity in a  square pulse with fixed time duration (10.55~$\mu$s for $^1$H and 9.45~$\mu$s for $^{13}$C nuclei). This allowed us to make finer adjustments to the $\alpha$ and $\theta$ values, as well as greater flexibility to explore the two scenarios aforementioned. We note that as the intensity of the rf-field is adjusted to set the time duration of the rotation pulses as of the order of tens of $\mu$s, the effects of the scalar coupling (with a frequency of order of hundreds of Hz) can be neglected during the single qubit rotation time. 

{\bf Wave and Particle Realism}

Wave and Particle Realism are obtained from a set of full QST of the bipartite system, which includes the controller and the qubit, parametrized by $\alpha$ and $\theta$. In each circuit depicted in Fig. \ref{Fig1}, to experimentally evaluate the degree of realism at the moment that the qubit is inside of the interferometer (after the phase shifter and before the last interference operation), we theoretically apply to the experimental state $\varrho_{exp}$ the maps: $\Phi_{W}(\varrho_{exp})=\sum_{a=-,+} (\Pi_{\mc{W}a}\otimes\mbb{1})\varrho_{exp}(\Pi_{\mc{W}a}\otimes \mbb{1})$, where $\Pi_{\mc{W}_{\pm}}=\ket{\mc{W}_{\pm}}\bra{\mc{W}_{\pm}}$ are the projectors for the wave observable, and $\Phi_{P}(\varrho_{exp})=\sum_{a=-,+} (\Pi_{\mc{P}a}\otimes\mbb{1})\varrho_{exp}(\Pi_{\mc{P}a}\otimes \mbb{1})$, with $\Pi_{\mc{P}_{\pm}}=\ket{\mc{P}_{\pm}}\bra{\mc{P}_{\pm}}$ being the projectors for the particle observable. We then compute the eigenvalues associated with these states and the corresponding entropies. The data displayed in Fig. \ref{Fig2} correspond to wave and particle realism computed as
\begin{equation}
    \mf{R}_W(\rho):=1+S(\varrho_{exp})-S\left(\Phi_W(\varrho_{exp})\right)
    \label{eq.10},
\end{equation}
and
\begin{equation}
    \mf{R}_P(\rho):=1+S(\varrho_{exp})-S\left(\Phi_P(\varrho_{exp})\right),
    \label{eq.11}
\end{equation}
parametrically plotted with respect to the measured visibility.

From the results presented in Figs. \ref{Fig2} and \ref{Fig3}{\bf c} we can conclude that in the proposed QCRE scenario (Fig.~\ref{Fig1}{\bf b}), the wave- or particle-like behaviour is firmly linked to the degree of definiteness of the quantum controller in the experiment preparation. The causal correlations between the qubit and the controller are established at an early stage, so that the notion of reality control is defensible. In this sense, we can claim that one does have a genuine ``controller'' in the experiment. This is a fundamental difference to all previous formulations of the QDCE (Fig.~\ref{Fig1}{\bf a}), wherein the quantum correlations between the qubit and the controller are established only after the qubit has travelled the interferometer. This requires an odd retrocausal picture where the future state of the controller is presumed to control the qubit reality in its past, even being uncorrelated with it at that moment. It is debatable, to say the least, whether one can think of some physical influence in this scenario.

{\bf  Complementarity and the measurement problem}

To further emphasise the adequacy of our approach in accounting for the physical reality and the complementarity principle, it is opportune to discuss another tricky facet of the wave-particle duality. In reference to the celebrated double-slit experiment, it is often pointed out that the system passes through the slits as a wave but clicks in a single well localised detector in each run of the experiment as a particle. It is clear that the pattern visibility built out of the many runs of the experiment has little to say about this conundrum. Moreover, one can readily recognise here, in connection with the wave function collapse, the conceptual difficulties associated with the measurement problem~\cite{schlosshauer2005decoherence}. A key point for the proper account of the issue is the realisation that measurements involve objects (detectors) whose physical role cannot be excluded from the system.

To make our discussion simple, we confine our attention to the open configuration of the QDCE. Let us take $\ket{\mf{r}_k}$ ($\mf{r}\equiv \tx{ready}$) and $\ket{\mf{a}_k}$ ($\mf{a}\equiv\tx{activated}$) as orthogonal internal states of the $k$-th detector. Here we consider the von-Neumann measurement model, which includes the measurement apparatus as a piece of ``the whole''. Before reaching the detectors, at some instant $t_i$, the state of the composite system reads
\begin{equation}
\ket{\psi_i}=\ket{\tx{out}}\left(\frac{e^{i\theta}\ket{0}+\ket{1}}{\sqrt{2}}\right)\ket{\mf{r}_0}\ket{\mf{r}_1}.
\label{psi_i}
\end{equation}
Applying our formalism yields $\mf{R}_P(\psi_i)=0$ and $\mf{R}_W(\psi_i)=1$, attesting the wave reality of the qubit during its flight inside the interferometer. This is the picture we can construct from the context implied by $\ket{\psi_i}$ at the instant $t_i$. After the interaction with the internal degrees of freedom of the detectors takes place, at some time $t_f$, the state of the joint system becomes
\begin{equation}
\ket{\psi_f}=\ket{\tx{out}}\left(\frac{e^{i\theta}\ket{0}\ket{\mf{a}_0}\ket{\mf{r}_1}+ \ket{1}\ket{\mf{r}_0}\ket{\mf{a}_1}}{\sqrt{2}}\right).
\label{psi_f}
\end{equation}

Now comes the crux: as recognised in Ref.~\cite{Dieguez18}, in all measurement processes the physical quantity to be observed is never directly accessed and actually is discarded. In fact, we just read the degree of freedom (of the detection apparatus) which has got quantum correlated with the desired quantity. The apparatus also possesses an irreducibly classical characteristic: it is rigidly attached to the reference frame, thus having well defined position and (null) velocity. To some extent, one may claim that this aspect makes direct contact with Bohr's view that ``any measurement must be essentially framed in terms of classical physical theories'' \cite{Camilleria2015}. For instance, in the Stern-Gerlach setup, we read the particle position (``the apppratus'') to get to know about the spin (the discarded degree of freedom). In the present instance, we can trace the qubit path out of the state \eqref{psi_f}, since this degree of freedom is never effectively accessed in the experiment. Indeed, the controller as well can be traced out because it also is out of the context imposed on us in the measurement process. This gives
\begin{equation}
 \varsigma=\tfrac{1}{2}\Big(\ket{\mf{a}_0}\bra{\mf{a}_0}\otimes\ket{\mf{r}_1}\bra{\mf{r}_1}+\ket{\mf{r}_0}\bra{\mf{r}_0}\otimes\ket{\mf{a}_1}\bra{\mf{a}_1}\Big).
\end{equation}
If we adopt $\ket{\mc{P}^+_k}\equiv\ket{\mf{a}_k}$ and $\ket{\mc{P}^-_k}\equiv\ket{\mf{r}_k}$ as particle-like states for the $k$-th detector (expressing definiteness of the detector internal state) and $\ket{\mc{W}^\pm_k}\equiv\left(\ket{\mf{a}_k}\pm\ket{\mf{r}_k}\right)/\sqrt{2}$ as wave-like states, then we can show that 
\begin{equation}
\mf{R}_{P_k}(\varsigma)=1\quad \tx{and}\quad \mf{R}_{W_k}(\varsigma)=0,
\end{equation} 
meaning that the internal states of the detectors, for the accessible context, are elements of reality. This explains why we never find the detection system in a superposition of realities $\ket{\mf{a}_0}\ket{\mf{r}_1}$ and $\ket{\mf{r}_0}\ket{\mf{a}_1}$, and hence relaxes the alleged tension involving the wave-particle duality and the measurement process (we note that in the present experiment we did not projected the control ancillary system). Most importantly, the present discussion reinforces the conceptual advantage of electing relation \eqref{BohrCP} as a formal statement of Bohr's complementarity principle.

{\bf Error Analysis}

The main sources of experimental errors are small uncontrolled variations on the transverse rf-fields intensities, non-idealities in its time modulation, and tiny inhomogeneities in the longitudinal static field as well as in the gradient pulses. The process to estimate the error propagation is based on a Monte Carlo method, sampling deviations of the quantum state tomography (QST) data with a Gaussian distribution having widths determined by the variances corresponding to such data. These data give us the necessary information to estimate the standard deviation of the distribution of the values for relevant quantities displayed in the figures. The variances of the tomographic data are obtained by preparing the same state QST one hundred times and comparing it with the theoretical expectation. Such procedure makes these variances include random and systematic errors in both state preparation and data acquisition by QST. The error in each element of the density matrix estimated from this analysis is about $1\%$.

The Monte Carlo error estimation of the realism quantifiers, displayed in Fig.~\ref{Fig2}, takes as inputs the QST uncertainty distribution of $\rho$ to obtain the error of the quantities displayed in Eqs.~(\ref{eq.10}) and (\ref{eq.11}). The
desired quantity error is obtained by randomly gathering the uncertainty in the experimentally reconstructed $\rho$. This procedure is then repeated many times with
new random gatherings. The resulting uncertainty
distribution of the desired quantity is directly obtained from the many random numerical trials. We additionally bound the random simulated fluctuations to the physically possible values for the realism quantifiers.

The majority of the experimental parameters, such as pulses intensity, phase, time duration, gradient intensity and variation, free evolution interval, and field's homogeneity are calibrated and optimized to minimize errors.

\vspace{2mm}
{\noindent \bf{\large{Data availability}}} 

The raw data that support the findings of this study may be available from the corresponding author upon reasonable request.



\vspace{5mm}
{\noindent \Large \bf{Acknowledgments}} \\
The authors acknowledge the Brazilian funding agencies Coordena\c{c}\~ao de Aperfei\c{c}oamento de Pessoal de N\'ivel Superior-Brasil
(CAPES)--Finance Code 001 (grant 88887.354951/2019-00, P.R.D), FAPESP, CNPq (grant 309373/2020-4, R.M.A.), Universidade Federal do Paran\'{a} (Edital 007/2021- PESQUISA/PRPPG/UFPR Projeto FUNPAR 3527), and the National Institute for Science and Technology of Quantum Information (CNPq, INCT-IQ 465469/2014-0). The authors are grateful to the Multiuser Central Facilities (UFABC) for the experimental support. J.P.S.P. thanks support from Innovation, Science and
Economic Development Canada, the Government of Ontario, and CIFAR. P. R. Dieguez acknowledges support by the Foundation for Polish Science (IRAP project, ICTQT, contract no. MAB/ 2018/5, co-financed by EU within Smart Growth Operational Programme). 

\vspace{2mm}
{\noindent \Large \bf{Author contributions}} \\
All authors developed the concept and designed the experiment; J. R. G, J. P. S. P., and R. M. S. performed the experiment; P. R. D and R. M. A, contributed to the theory. All authors discussed the results, contributed to analyzing the data and writing the paper.

\vspace{2mm}
{\noindent \bf{\large{Competing interests}}}\\  
The authors declare no competing interests.


\end{document}